\documentclass[journal]{IEEEtran}
\usepackage[utf8]{inputenc}
\usepackage{stfloats}
\usepackage{cite}
\usepackage{amsmath}
\usepackage{amsfonts}
\usepackage{dsfont}
\usepackage{graphicx}
\usepackage{changepage}
\usepackage{color}
\usepackage{epstopdf}
\usepackage[utf8]{inputenc}
\UseRawInputEncoding

\ifCLASSINFOpdf

\fi

\begin{document}
\title{Performance Analysis of Dual-Hop Mixed PLC/RF Communication Systems}
\author{Liang~Yang,~Xiaoqin~Yan,~Sai~Li,~Daniel~Benevides~da~Costa,~and Mohamed-Slim~Alouini}
\maketitle
\begin{abstract}
In this paper, we study a dual-hop mixed power line communication and radio-frequency communication (PLC/RF) system, where the connection between the PLC link and the RF link is made by a decode-and-forward (DF) or amplify-and-forward (AF) relay. Assume that the PLC channel is affected by both additive background noise and impulsive noise suffers from Log-normal fading, while the RF link undergoes Rician fading. Based on this model, analytical expressions of the outage probability (OP), average bit error rate (BER), and the average channel capacity (ACC) are derived. Furthermore, an asymptotic analysis for the OP and average BER, as well as an upper bound expression for the ACC are presented. At last, numerical results are developed to validate our analytical results, and in-depth discussions are conducted.
\end{abstract}
\IEEEpeerreviewmaketitle
\section{Introduction}
\IEEEPARstart{A}{s} a low-cost and energy-saving communication technology, power line communication (PLC) utilizes the existing cables for data transmission \cite{1}-\cite{3}. According to different voltage levels, PLC can communicate through low-voltage cables, medium-voltage cables, and high-voltage cables \cite{4}-\cite{5}. Compared with other methods of communication, PLC has the characteristics of wide coverage, convenient connection, and no need to rewire, which enables it to be used in indoor and outdoor communication. For instance, in \cite{6}, the authors put forward a kind of indoor narrow-band PLC network model, and provided the appropriate types of cables and electrical appliances through laboratory experiments and simulation results. Based on the deep integration of PLC and visible light (VLC), a original, practical, and economical indoor broadband broadcasting network was studied in \cite{7}. The authors in \cite{8} investigated the spatial correlation in indoor multiple-input multiple-output (MIMO) PLC channels. Additionally, PLC has arisen as one of the main technical methods of two-way communication in smart grid (SG) \cite{9}. It can be connected to all locations of the grid and transmit data on the original infrastructure. In \cite{10}, the authors studied the performance of discrete wavelet multitone transceiver for narrow-band PLC in SG.

However, compared with general wireless communication systems, the performance of PLC communication systems is highly influenced by frequency selectivity, path loss, and various attenuation and interference. In addition, impedance mismatch and non-Gaussian noise in the PLC channel arise as further issues to be tackled with. Moreover, since PLC was originally used to transmit electric energy (and not for data communication), the transmission power in PLC systems should comply with the regulations of relevant government departments, which leads to the limitation of system capacity and transmission distance \cite{11}. Finally, the affection of background noise (BGN) and impulsive noise (IMN) on system performance should also be jointly considered in PLC systems as they are the main reasons for data loss \cite{12}. In order to alleviate these inconveniences, researchers in the field of PLC have proposed to combine PLC with wireless communication technologies, such as multi-antenna schemes, cooperative communications, MAC protocols, and relaying methods \cite{13}-\cite{15}.

Along the years, relay technology has shown to improve the reliability and coverage of the system. Relying on this fact, relay-aided PLC systems have been widely investigated. In \cite{16}, a PLC system with Log-normal (LN) fading and Bernoulli-Gaussian IMN assisted by amplify-and-forward (AF) relay was studied, which was shown to outperform PLC systems without embedded relay. The outage probability (OP), average bit error rate (BER), and average channel capacity (ACC) of a mixed decode-and-forward (DF) relay-aided PLC system were derived in \cite{17}, while a full-duplex AF relay for PLC networks was considered in \cite{18}. Moreover, the performance of half-duplex PLC networks with either AF or DF relays was examined in \cite{19}. With the aim to enhance the energy efficiency of relaying PLC systems, AF relaying with energy-harvesting capabilities was embedded in a PLC system and accurate expressions for energy efficiency of such systems were derived in \cite{20}. In \cite{21}, the authors studied the physical layer security of the relay-aided PLC system with eavesdropping and noise interference. More recently, the authors considered a PLC system with incremental AF and DF relaying, and the results showed a great spectral efficiency improvement \cite{22}.

In the aforementioned works dealing with relay-aided PLC systems, the RF links were assumed to undergo LN or Rayleigh fading, although Rician fading arises as a more suitable model owing to the presence of the line-of-sight (LoS) component. As far as the authors know, the performance of PLC/RF systems with both BGN and IMN, where the respective links are modeled by a LN and a Rician distribution, respectively, has not been investigated in the field of PLC. Thus, this paper presents and studies a dual-hop mixed PLC/RF system, assuming both DF and AF relaying protocols.

The original contributions of this paper can be stated as follows:
\begin{itemize}
\item[$\bullet$] A cost-effective and efficient PLC/RF system is introduced and analyzed. The RF system and the PLC system are connected by both DF and AF relaying protocols. The channel of the PLC link is modeled by LN fading and it is subject to additive BGN and IMN, while the RF channel follows a Rician fading distribution.
\item[$\bullet$] For DF relaying, respective expressions for the OP, average BER, and ACC are derived. To get further insights, asymptotic analysises for the OP and average BER are carried out. Moreover, we also derive the upper bound expression for ACC.
\item[$\bullet$] Novel closed-form expressions for the cumulative distribution function (CDF) and probability density function (PDF) of the end-to-end signal-to-noise (SNR) of the considered system with an AF relay are obtained. Moreover, the respective expressions for OP, average BER, as well as ACC are achieved.
\item[$\bullet$] The influences of critical system parameters on the overall performance is studied and insightful discussions are drawn.
\item[$\bullet$] The Monte Carlo simulation results validate our analytical results.
\end{itemize}
\begin{figure}[t]
 \centering
 \includegraphics[scale=0.5]{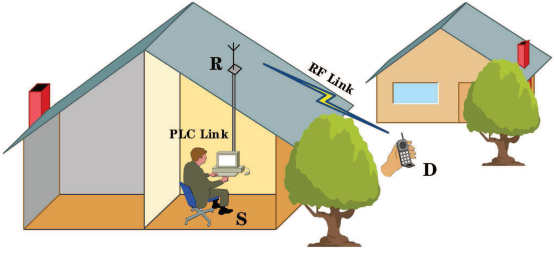}
 \caption{System setup.}
\end{figure}
The remaining section of this paper is arranged as follows. Section \uppercase\expandafter{\romannumeral 2}
introduces the system and channel models. In Section \uppercase\expandafter{\romannumeral 3}, a comprehensive performance of the considered system is analyzed, including OP, average BER, and ACC. Section \uppercase\expandafter{\romannumeral 4} provides illustrative numerical examples along with insightful discussions. Finally, Section \uppercase\expandafter{\romannumeral 5} concludes the paper. Appendix A provides a detailed proof for the CDF and PDF of the end-to-end SNR.

\section{System and Channel Models}
Consider a dual-hop mixed PLC/RF communication system, which includes a source node (S), a relay node (R) and a destination node (D), as shown in Fig. 1. In the first time-slot $T_{1}$, the source S transmits data to R through the PLC link modeled by a LN fading distribution with BGN and IMN. Then, during the second time-slot $T_{2}$, by employing DF or AF relaying protocols at R, the signals are sent to D over an RF link modeled by Rician fading. Suppose that direct connection between S and D does not exist.
\subsection{The PLC Link}
By utilizing a binary modulation scheme, the symbol $x$ from S is transmitted to R through the power cables. Thus, the signal at R can be presented as
\begin{equation}
y_{\text{SR}} = h_{\text{SR}}x+n_{\text{SR}},
\end{equation}
where $n_{\text{SR}}$ represents the
additive noise of the PLC channel and $h_{\text{SR}}$ denotes the channel fading factor. From \cite{23}, $h_{\text{SR}}$ is modeled by a LN distribution and its PDF can be expressed as
\begin{equation}
f_{h_{\text{SR}}}(h_{\text{SR}})=\frac{1}{h_{\text{SR}}\sqrt{2\pi \sigma _{\text{SR}}^{2}}}{\rm exp}\left ( {-}\frac{\left (\ln(h_{\text{SR}})-\mu _{\text{SR}}\right )^{2}}{2\sigma _{\text{SR}}^{2}} \right ),
\end{equation}
where $\sigma_{\text{SR}}^{2}$ and $\mu_{\text{SR}}$ denote the variance and mean of $\ln(h_{\text{SR}})$,
respectively. Due to the random transient switching of low-power components and electrical equipment connected
to cables in PLC systems, the interference of both BGN and IMN are considered. In this case, Poisson$-$Gaussian mixture statistical \cite{23} is employed to model the noise. Thus, the noise in the PLC link can be represented as $n_{\text{SR}}= n_{b}{+}n_{i}n_{p}$, where $n_{b}$
is the BGN modeled as the additive white Gaussian noise (AWGN) with zero mean and variance
$\sigma_{b}^{2}$, $n_{i}n_{p}$ denotes the IMN occurring during $T$, with $n_{p}$
being defined as the occurrence of the IMN with a rate of $\lambda$ units per second in the system,
which is modeled by a Poisson process, and $n_{i}$ represents the AWGN with mean zero and
variance $\sigma_{i}^{2}$. Assuming only the real part of the noise $n_{\text{SR}}$, its PDF can
be written as \cite{23}
\begin{align}
f_{n_{\text{SR}}}(n_{\text{SR}})=&\frac{1{-}P_{i}}{\sqrt{2\pi \sigma _{b}^{2}}}{\rm exp}\left ( {-}\frac{n_{\text{SR}}^{2}}{2\sigma _{b}^{2}} \right ){+}\frac{P_{i}}{\sqrt{2\pi \sigma _{b}^{2}\left ( 1{+}\eta  \right )}}\nonumber\\
&\times {\rm exp}\left ( {-}\frac{n_{\text{SR}}^{2}}{2\sigma _{b}^{2}\left ( 1{+}\eta  \right )} \right ),
\end{align}
where $P_{i}=\lambda T$ represents the probability of the occurrence of impulsive noise, and $\eta=\sigma_{i}^{2}/\sigma_{b}^{2}$ denotes the ratio of the powers of IMN to BGN.

\begin{figure*}[b]
\hrulefill
\begin{align*}
&f_{\gamma _{o}}\left ( \gamma  \right ) =
\frac{P_{i}m_{2}\exp(-{m_{2}\over\Omega_{2}}\gamma)}{\Omega_{2}\Gamma(m_2)\exp(K)}\sum_{l=0}^{k}\sum_{r=0}^{m_{2}}
\binom{m_{2}}{r}\left({m_{2}\gamma\over \Omega_{2}}\right)^{l{+}m_{2}{-}r}
{k^{1{-}2l}K^l(C(K{+}1))^{l{+}1}\Gamma(k{+}l)\over \overline{\gamma}_{RD}^{l{+}1}\Gamma(k{-}l{+}1)\Gamma^2(l{+}1)}
G_{0,2}^{2,0}\left[\left.\frac{Cm_{2}(1{+}K)\gamma}{\Omega_{2}\overline{\gamma}_{RD}}\right|\begin{matrix}{-}\\{r{-}l{-}1,0}\end{matrix}\right]\nonumber\\
&{+}\frac{\left(1{-}P_{i}\right)m_{1}\exp(-{m_{1}\over\Omega_{1}}\gamma)}{\Omega_{1}\Gamma(m_1)\exp(K)}\sum_{l=0}^{k}\sum_{r=0}^{m_{1}}
\binom{m_{1}}{r}\left({m_{1}\gamma\over \Omega_{1}}\right)^{l{+}m_{1}{-}r}
{k^{1{-}2l}K^l(C(K{+}1))^{l{+}1}\Gamma(k{+}l)\over \overline{\gamma}_{RD}^{l{+}1}\Gamma(k{-}l{+}1)\Gamma^2(l{+}1)}
G_{0,2}^{2,0}\left[\left.{Cm_{1}(1{+}K)\gamma\over\Omega_{1}\overline{\gamma}_{RD}}\right|\begin{matrix}{-}\\{r{-}l{-}1,0}\end{matrix}\right]\nonumber.
\tag{16}
\end{align*}
\end{figure*}
\begin{figure*}[b]
\hrulefill
\begin{align*}
F_{\gamma _{o}}\left ( \gamma  \right ) = & F_{\gamma_{SR}}(\gamma){+}\frac{\left(1{-}P_{i}\right)m_{1}\exp({-}{m_{1}\over\Omega_{1}}\gamma)}{\Omega_{1}\Gamma(m_1)\exp(K)}\sum_{l=0}^{k}\sum_{r=0}^{m_{1}{-}1}
\binom{m_{1}{-}1}{r}\left({m_{1}\gamma\over \Omega_{1}}\right)^{m_{1}{-}r{-}1}B_{l}
G_{1,3}^{2,1}\left[\left.\frac{Cm_{1}(1{+}K)\gamma}{\Omega_{1}\overline{\gamma}_{RD}}\right|\begin{matrix}{1}\\{1{+}r,1{+}l,0}\end{matrix}\right]\nonumber\\
&+\frac{P_{i}m_{2}\exp({-}{m_{2}\over\Omega_{2}}\gamma)}{\Omega_{2}\Gamma(m_2)\exp(K)}\sum_{l=0}^{k}\sum_{r=0}^{m_{2}{-}1}
\binom{m_{2}{-}1}{r}\left({m_{2}\gamma\over \Omega_{2}}\right)^{m_{2}{-}r{-}1}B_{l}
G_{1,3}^{2,1}\left[\left.\frac{Cm_{2}(1+K)\gamma}{\Omega_{2}\overline{\gamma}_{RD}}\right|\begin{matrix}{1}\\{1{+}r,1{+}l,0}\end{matrix}\right].
\tag{17}
\end{align*}
\end{figure*}

When there are only BGN samples in the PLC channel, the resulting SNR of the PLC link can be written as
\begin{equation}
\gamma_{\text{SR}1}=\frac{E_{b}\left | h_{\text{SR}} \right|^{2}}{\sigma_{b}^{2}}=\overline{\gamma}_{\text{SR}1}\left|h_{\text{SR}}\right|^{2},
\end{equation}
where $\overline{\gamma}_{\text{SR}1}$ denotes the average SNR of the first hop when
only BGN samples are presented and $E_{b}$ represents the
average energy of the signal. Similarly, when the IMN
samples and BGN samples appear simultaneously in
the PLC link, the instantaneous SNR can be written as
\begin{equation}
\gamma _{\text{SR}2}=\frac{E_{b}\left | h_{\text{SR}} \right |^{2}}{\sigma _{b}^{2}\left ( 1{+}\eta  \right )}=\overline{\gamma }_{\text{SR}2}\left | h_{\text{SR}} \right |^{2},
\end{equation}
where $\overline{\gamma }_{\text{SR}2}$ is the average SNR of the PLC link
when IMN samples and BGN samples occur simultaneously in the system.
According to \cite{23}, the PDF of the SNR $\gamma _{\text{SR}}$ of the PLC link can be shown to be given by
\begin{align}
f_{\gamma _{\text{SR}}}\left ( \gamma  \right )=&\left ( 1{-}P_{i} \right )\left ( \frac{m_{1}}{\Omega_{1}} \right )^{m_{1}}\frac{\gamma ^{m_{1}{-}1}}{\Gamma \left ( m_{1} \right )}{\rm exp}\left ( {-}\frac{m_{1}}{\Omega _{1}} \gamma \right )\nonumber\\
&{+}P_{i}\left ( \frac{m_{2}}{\Omega _{2}} \right )^{m_{2}}\frac{\gamma ^{m_{2}{-}1}}{\Gamma \left ( m_{2} \right )}{\rm exp}\left ( {-}\frac{m_{2}}{\Omega _{2}}\gamma  \right ),
\end{align}
where $m_{1}$ and $m_{2}$
are defined as the shadowing severity parameters of the Gamma PDF,
$\Omega _{1}$ and $\Omega _{2}$ represent the shadowing area mean power of the Gamma PDF, and $\Gamma(\cdot)$ is the Gamma function \cite{24}. Thus, the CDF of $\gamma_{\text{SR}}$ can be represented as \cite{23}
\begin{align}
F_{\gamma _{\text{SR}}}\left ( \gamma  \right )=&
\frac{1{-}P_{i}}{\Gamma \left ( m_{1} \right )}
G_{1,2}^{1,1}\left[\left.\frac{m_{1} }{\Omega_{1}}\gamma \right|\begin{matrix}1\\m_{1},0\end{matrix}\right]\nonumber\\
&{+}\frac{P_{i}}{\Gamma \left ( m_{2} \right )}
G_{1,2}^{1,1}\left[\left.\frac{m_{2} }{\Omega_{2}}\gamma \right|\begin{matrix}1\\m_{2},0\end{matrix}\right],
\end{align}
where $G_{c,d}^{m,n}\left [ \cdot \right]$ is the Meijer $G$-function \cite{24}.
\subsection{The RF Link}
In $T_{2}$, R uses DF or AF relaying protocols to forward the received signals to D via the RF channel. It is worthy to say that
the RF link between R and D is considered to follow a Rician fading distribution.
\subsubsection{DF Case}
For the DF case, we can write the received signal at D as
\begin{equation}
y_{\text{RD}}^{DF} = \sqrt{P_{R}}h_{\text{RD}}\widehat{x} + n_{\text{RD}},
\end{equation}
where $P_{R}$ denotes the average transmit power at D, $\widehat{x}$ is defined as the signal transmitted from R, $n_{\text{RD}}$ represents the AWGN term with zero mean and variance $N$, and $h_{\text{RD}}$ denotes the RF channel coefficient. From (8), the resulting SNR of the RF
link can be formulated as
\begin{align}
\gamma_{\text{RD}}{=}\frac{P_{R}\left |h_{RD} \right|^{2}}{N }{=}\overline{\gamma}_{\text{RD}}\left |h_{RD} \right|^{2},
\end{align}
where $\overline{\gamma}_{RD}$ is the average SNR of the RF link.
The PDF and CDF of $\gamma_{\text{RD}}$ are given by \cite{25}
\begin{align}
f_{\gamma _{RD}} (\gamma ){=}&\frac{K{+}1}{\overline{\gamma}_{RD}}\exp \left({-}\frac{(K{+}1)\gamma_{RD}}{\overline{\gamma}_{RD}}{-}K \right)\nonumber\\
&\times I_{0}\left(2\sqrt{K(K{+}1)\frac{
\gamma_{RD}}{\overline{\gamma}_{RD}}} \right),
\end{align}
and
\begin{align}
F_{\gamma _{RD}}\left(\gamma \right)=1-Q_{1}\Bigg (\sqrt{2K}, \sqrt{\frac{2\left(K+1 \right)}{\overline{\gamma}_{RD}}\gamma} \Bigg),
\end{align}
where $I_{0}(\cdot)$ is the zeroth-order modified Bessel function of the first kind \cite{24}, $K$ $(K\geq 0)$ represents the Rician factor, and $Q_{1}(\cdot,\cdot)$ is the Marcum Q-function of the first order \cite{26}.

\subsubsection{AF Case}
For the AF case, the received signal at D is
\begin{align}
y_{\text{RD}}^{AF} = \sqrt{P_{R}}Gh_{\text{RD}}y_{\text{SR}}+n_{\text{RD}},
\end{align}
where $G$ is the fixed amplifying gain.
Thus, the overall instantaneous SNR can be expressed as
\begin{align}
\gamma_{o} = \frac{\gamma_{\text{SR}}\gamma_{\text{RD}}}{C+\gamma_{\text{RD}}},
\end{align}
where $C$ is a constant determined by the AF relay gain $G$.

To derive the expression of the PDF of the overall instantaneous SNR $\gamma_{o}$ and facilitate our calculation, the approximation of $I_{v}(x)\simeq\sum_{l=0}^{k}{\Gamma(k+l)\over \Gamma(l+1)\Gamma(k-l+1)}{k^{1-2l}\over \Gamma(v+l+1)}\left({x\over 2}\right)^{v+2l}$ \cite{26} is used in (10) for $0<x<2k$.
Thus (10) and (11) can be rewritten as
\begin{align}
&f_{\gamma _{RD}} (\gamma ){\simeq}\frac{K{+}1}{\overline{\gamma}_{RD}}\exp \left({-}\frac{(K{+}1)\gamma_{RD}}{\overline{\gamma}_{RD}}{-}K \right)\nonumber\\
&\times \sum_{l=0}^{k}{\Gamma(k+l)k^{1-2l}K^l(K+1)^l\over \Gamma^2(l+1)\Gamma(k-l+1)\overline{\gamma}_{RD}^l}\gamma^{l},
\end{align}
and
\begin{align}
&F_{\gamma_{RD}}(\gamma){\simeq}1{-}\exp \left({-}\frac{(K{+}1)\gamma_{RD}}{\overline{\gamma}_{RD}}{-}K \right)\nonumber\\
&\times\sum_{l=0}^{k}\sum_{r=0}^{l}{k^{1{-}2l}K^l(K{+}1)^r\Gamma(l{+}k)\over \Gamma(r{+}1)\Gamma^2(l{+}1)\Gamma(k{-}l{+}1)\overline{\gamma}_{RD}^r}\gamma^r.
\end{align}
Then, using the similar method \cite{27}, the PDF and CDF of $\gamma_{o}$ are derived in (16) and (17) when $m_{1}$ and $m_{2}$ are integers, shown at the bottom of this page, where $B_{l}{=}{k^{1-2l}K^l\Gamma(k+l)\over \Gamma(k-l+1)\Gamma^{2}(l+1)}$.

\emph{Proof:} See Appendix \uppercase\expandafter{A}.

\section{Performance Analysis}
\subsection{DF Relaying}
\subsubsection{OP}
The OP can be defined as the probability that the
overall instantaneous SNR is lower than a certain SNR threshold $\gamma_{\text{th}}$.
Thus, by relying on (7) and (11), and making the appropriate substitutions, the outage probability can be determined from the expression below
\begin{align}
P_{\text{out}}^{DF} &=\text{Pr}\left(\text{min}(\gamma _{\text{SR}},\gamma _{\text{RD}}){<} \gamma_{\text{th}} \right) \nonumber\\
&=F_{\gamma_{\text{SR}}}(\gamma_{\text{th}}){+}F_{\gamma_{\text{RD}}}(\gamma_{\text{th}}){-}F_{\gamma_{\text{SR}}}(\gamma_{\text{th}})F_{\gamma_{\text{RD}}}(\gamma_{\text{th}}).\nonumber
\tag{18}
\end{align}
In order to gain further insights on outage probability with the DF protocol, next we derive an asymptotic outage expression.
At high SNR, the last term of (18) can be ignored. Then, making use of the following asymptotic series expansion
of the Meijer G-function [28, Eq. (07.34.06.0040.01)]
\begin{align}&G_{c,d}^{m,n}\left(z\Big \vert _{b_1,...,b_d}^{a_1,...,a_c}\right)\nonumber\\&=\sum _{\iota=1}^m \displaystyle \frac{\prod _{j=1,j\ne \iota}^m\Gamma (b_j{-}b_\iota)\prod _{j=1}^n\Gamma (1{-}a_j{+}b_\iota)}{\prod _{j=n+1}^c\Gamma (a_j{-}b_\iota)\prod _{j=m+1}^d\Gamma (1{-}b_j{+}b_\iota)}z^{b_\iota}(1{+}o(z)), \nonumber
\tag{19}
\end{align}
the asymptotic $F_{\gamma_{SR}}(\gamma)$ can be represented as
\begin{align}
\underset{\overline{\gamma}_{SR}\rightarrow \infty}{F_{\gamma_{SR}}(\gamma)}{\simeq}\frac{1{-}P_{i}}{\Gamma(1{+}m_{1})}\left(\frac{m_{1}\gamma}{\Omega_{1}}\right)^{m_{1}}{+}\frac{P_{i}}{\Gamma \left( 1{+}m_{2}\right)}\left(\frac{m_{2}\gamma}{\Omega_{2}}\right)^{m_{2}}.\nonumber
\tag{20}
\end{align}
Furthermore, the asymptotic expression for $F_{\gamma_{RD}}(\gamma)$ can be expressed as \cite{29}
\begin{align}
\underset{\overline{\gamma}_{RD}\rightarrow \infty}{F_{\gamma_{RD}}(\gamma)}{\simeq}{\left(1{+}K\right)\gamma\over \overline{\gamma}_{RD}\exp \left(K \right)}.\nonumber\
\tag{21}
\end{align}
Finally, the asymptotic outage probability is attained from the sum of (20) and (21) by setting $\gamma=\gamma_{\text{th}}$.
\subsubsection{Average BER}
Generally, the average BER of a DF relaying system can be formulated as
\begin{align}
P_{\text{BER}}^{DF}{=}P_{e1}{+}P_{e2}{-}2P_{e1}P_{e2},\nonumber
\tag{22}
\end{align}
where $P_{e1}$ and $P_{e2}$ are the average BER of the first hop and the second hop,
respectively. In addition, the average BER for various binary modulations can be written as \cite{30}
\begin{align}
P_{b}{=}\frac{q^{p}}{2\Gamma(p)}\int_{0}^{\infty}{\rm exp}(-q\gamma)\gamma^{q{-}1}F_{\gamma}(\gamma)d\gamma,\nonumber
\tag{23}
\end{align}
where $p$ and $q$ are parameters related to modulation schemes. In our analysis, we consider the differential binary phase shift keying (DBPSK) scheme (i.e., $p=1, q=1$). Therefore, $P_{1}$ and $P_{2}$ can be
derived as
\begin{align}
P_{e1}{=}\frac{1{-}P_{i}}{2\Gamma \left ( m_{1} \right)}G_{2,2}^{1,2}\left[\left.\frac{m_{1} }{\Omega_{1}} \right|\begin{matrix}0,1\\m_{1},0\end{matrix}\right]
{+}\frac{P_{i}}{2\Gamma(m_2)}G_{2,2}^{1,2}\left[\left.\frac{m_{2} }{\Omega_{2}} \right|\begin{matrix}0,1\\m_{2},0\end{matrix}\right],\nonumber
\tag{24}
\end{align}
\begin{align}
P_{e2}{=}{1{+}K\over 2\left(1{+}K{+}\overline{\gamma}_{RD}\right)\exp(K)}{}_{1}F_{1}\left(1;1;{K\left(K{+}1\right)\over K{+}1{+}\overline{\gamma}_{RD}}\right),
\nonumber
\tag{25}
\end{align}
where ${}_{1}F_{1}\left(\cdot\right)$ is the confluent hypergeometric function \cite{24}.
By substituting (24) and (25) into (22), we obtain the average BER expression.
Since the last term in (22) can be negligible at high SNRs,
one can obtain the asymptotic average BER as follows
\begin{align}
P_{\text{BER}}^{DF}\rightarrow P_{e1}^{A}+P_{e2}^{A},\nonumber
\tag{26}
\end{align}
where $P_{e1}^{A}$ and $P_{e2}^{A}$ denote the asymptotic BER of $P_{e1}$ and $P_{e2}$, respectively. Substituting (20) into (23), the expression of $P_{e1}^A$ can be obtained by
\begin{align}
P_{e1}^A={1{-}P_{i}\over 2}\left({m_{1} \over \Omega_{1} }\right)^{m_{1}}{+}{ P_{i}\over 2 }\left({m_{2}\over \Omega_{2}}\right)^{m_{2}}.\nonumber
\tag{27}
\end{align}
Then, according \cite{29}, the $P_{e2}^{A}$ can be expressed as
\begin{align}
P_{e2}^{A}=\frac{\left(1{+}K\right)\Gamma(2)}{ 2\overline{\gamma}_{RD}\exp(K)}.\nonumber
\tag{28}
\end{align}

\begin{figure*}[t]
\begin{align*}
\mathbb{E}(\gamma)&{=}{\left(1{-}P_{i}\right)\left( { m_{1}\over \Omega_{1} }\right)^{m_{1}}\over \Gamma(m_{1})\exp(K)}
\sum_{l=0}^{k}\sum_{r=0}^{l}\Gamma(l+1)\Gamma(m_{1}{+}r)B_{l}\Theta_{r}W_{1r}
{+}{P_{i}\left( { m_{2}\over \Omega_{2} }\right)^{m_{2}}\over \Gamma(m_{2})\exp(K)}
\sum_{l=0}^{k}\sum_{r=0}^{l}\Gamma(l+1)\Gamma(m_{2}{+}r)B_{l}\Theta_{r}W_{2r}\nonumber\\
&{-}{\left(1{-}P_{i}\right)\overline{\gamma}_{RD}\exp({-}K)\over \left(K{+}1\right)\Gamma(m_{1})}\sum_{l=0}^{k}B_{l}
G_{2,2}^{1,2}\left[\left. {m_{1}\overline{\gamma}_{RD}\over \Omega_{1}\left(K{+}1\right)} \right|\begin{matrix}{-}l{-}1,1\\ m_{1},0\end{matrix} \right]
-{P_{i}\overline{\gamma}_{RD}\exp(-K)\over \left(K{+}1\right)\Gamma(m_{2})}\sum_{l=0}^{k}B_{l}
G_{2,2}^{1,2}\left[\left. {m_{2}\overline{\gamma}_{RD}\over \Omega_{2}\left(K{+}1\right)} \right|\begin{matrix}{-}l{-}1,1\\ m_{2},0\end{matrix} \right]\nonumber\\
&{+}{\overline{\gamma}_{RD} \over (K{+}1)\exp(K)}\sum_{l=0}^{k}B_{l}\Gamma(l{+}2).
\nonumber
\tag{35}
\end{align*}
\end{figure*}

\begin{figure*}[t]
\hrulefill
\begin{align*}
P_{\text{BER}}^{AF} {=}&\frac{1{-}P_{i}}{2\Gamma \left ( m_{1} \right )}G_{2,2}^{1,2}\left[\left.\frac{m_{1} }{\Omega_{1}} \right|\begin{matrix}0,1\\m_{1},0\end{matrix}\right]{+}\frac{P_{i}}{2\Gamma \left ( m_{2} \right )}G_{2,2}^{1,2}\left[\left.\frac{m_{2} }{\Omega_{2}} \right|\begin{matrix}0,1\\m_{2},0\end{matrix}\right]\nonumber \\
&{+}\frac{(1{-}P_{i})\Omega_{1}\exp({-}K)}{2\Gamma(m_{1}{+}1)}\sum_{l=0}^{k}\sum_{r=0}^{m_{1}{-}1}\begin{pmatrix}{m_{1}{-}1}\\{r} \end{pmatrix}\left ( \frac{m_{1}{+}\Omega_{1}}{m _{1}} \right )^{r{-}m_{1}}B_{l}
G_{2,3}^{2,2}\left[ \left. { Cm_{1}(1{+}K) \over \overline{\gamma}_{RD}(m_{1}{+}\Omega_{1})} \right|{}_{1{+}r,1{+}l,0}^{1{+}r{-}m_{1},1} \right ]\nonumber\\
&{+}\frac{P_{i}\Omega_{2}\exp({-}K)}{2\Gamma(m_{2}{+}1)}\sum_{l=0}^{k}\sum_{r=0}^{m_{2}{-}1}\binom{m_{2}{-}1}{r}\left ( \frac{m_{2}{+}\Omega_{2}}{m _{2}} \right )^{r{-}m_{2}}B_{l}
G_{2,3}^{2,2}\left[ \left. { Cm_{2}(1{+}K) \over \overline{\gamma}_{RD}(m_{2}{+}\Omega_{2})} \right|{}_{1{+}r,1{+}l,0}^{1{+}r{-}m_{2},1} \right ].\nonumber
\tag{37}
\end{align*}
\end{figure*}
\begin{figure*}[t]
\hrulefill
\begin{align*}
C_{AF}=
&\frac{(1{-}P_{i})\exp(-K)}{2\Gamma(m_{1})\ln(2)}\sum_{l=0}^{k}\sum_{r=0}^{m_{1}}\begin{pmatrix}
m_{1}\\r \end{pmatrix}\left({C(1{+}K)\over \overline{\gamma}_{RD}}\right)^{l{+}1}B_{l}
G_{1,0;2,2;;0,2}^{0,1;1,2;2,0}\left[ \left. \begin{matrix} r{-}m_{1}-l \\ - \end{matrix} \right|
\left. \begin{matrix} 1,1 \\ 1, 0 \end{matrix} \right|
\left. \begin{matrix}  - \\ r-l-1,0\end{matrix}\right|
\frac{\Omega_{1}}{m_{1}},\frac{C(1+K)}{\overline{\gamma}_{RD}}\right ]\nonumber\\
&{+}\frac{P_{i}\exp(-K)}{2\Gamma(m_{2})\ln(2)}\sum_{l=0}^{k}\sum_{r=0}^{m_{2}}\begin{pmatrix}
m_{2}\\ r \end{pmatrix}\left({C(1{+}K)\over \overline{\gamma}_{RD}}\right)^{l{+}1}B_{l}
G_{1,0;2,2;;0,2}^{0,1;1,2;2,0}\left[ \left. \begin{matrix} r{-}m_{2}-l \\ - \end{matrix} \right|
\left. \begin{matrix} 1,1 \\ 1, 0 \end{matrix} \right|
\left. \begin{matrix}  - \\ r-l-1,0\end{matrix}\right|
\frac{\Omega_{2}}{m_{2}},\frac{C(1+K)}{\overline{\gamma}_{RD}}\right ].\nonumber
\tag{39}
\end{align*}
\hrulefill
\end{figure*}

\subsubsection{Average Channel Capacity}
Generally, the overall capacity of the dual-hop DF system can be defined as
\begin{align}
C_{DF}&{=}\frac{1}{2}\mathbb{E}\left\{\text{log}_{2}(1{+}\gamma) \right\}
{=}\frac{1}{2\text{ln}(2)}\int_{0}^{\infty}\text{ln}(1{+}\gamma)f_{\gamma}(\gamma)d\gamma \nonumber\\
&{=}\frac{1}{2\text{ln}(2)}(C_{1}{+}C_{2}{-}C_{3}-C_{4}),\nonumber
\tag{29}
\end{align}
where $f_{\gamma}(\gamma)=f_{\gamma_{SR}}(\gamma)+f_{\gamma_{RD}}(\gamma)-f_{\gamma_{SR}}(\gamma)F_{\gamma_{RD}}(\gamma)-F_{\gamma_{SR}}(\gamma)f_{\gamma_{RD}}(\gamma)$.
By utilizing ${\rm ln}(1+\gamma){=}G_{2,2}^{1,2}\left[\left.\gamma \right|{}_{1,0}^{1,1}\right]$
[28, Eq. (01.04.26.0003.01)], $\exp({-}bz){=}G_{0,1}^{1,0}\left[\left.bz \right|{}_{0}^{-}\right]$ [28, Eq. (07.34.21.0013.01)], [28, Eq. 07.34.21.0088.01] and [28, Eq. 07.34.21.0081.01],
expressions for $C_{1}$, $C_{2}$, $C_{3}$ and $C_{4}$ can be calculated as
\begin{align}
C_{1}
{=}&\frac{1{-}P_{i}}{\Gamma\left(m_1\right)}
G_{3,2}^{1,3}\left[\left. \frac{\Omega_{1}}{m_{1}} \right|\begin{matrix} 1{-}m_{1},1,1\\1,0 \end{matrix} \right]\nonumber\\
&+\frac{P_{i}}{\Gamma\left(m_{2}\right)}
G_{3,2}^{1,3}\left[\left. \frac{\Omega_{2}}{m_{2}} \right|\begin{matrix} 1{-}m_{2},1,1\\1,0 \end{matrix} \right],\nonumber
\tag{30}
\end{align}
\begin{align}
C_{2}
{=}&\sum_{l=0}^{k}{\Gamma(k{+}l)k^{1{-}2l}K^l \over \Gamma(k{-}l+1)\Gamma^2(l{+}1)\exp(K)}
G_{3,2}^{1,3}\left[\left. \frac{\overline{\gamma}_{RD}}{K{+}1} \right|\begin{matrix} {-}l,1,1\\1,0 \end{matrix} \right],\nonumber
\tag{31}
\end{align}
\begin{align}
&C_{3}
{=}{(P_{i}{-}1)\left( { m_{1}\over \Omega_{1} }\right)^{m_{1}}\over \Gamma(m_{1})\exp(K)}\sum_{l=0}^{k}\sum_{r=0}^{l}\Gamma(l{+}1)B_{l}\Theta_{r}
W_{1r}G_{3,2}^{1,3}\left[\left. W_{1r} \right|{}_{ \varsigma_{1r}}^{1,0} \right] \nonumber\\
&{+}C_{1}{-}{P_{i}\left( { m_{2}\over \Omega_{2} }\right)^{m_{2}}\over \Gamma(m_{2})\exp(K)}
\sum_{l=0}^{k}\sum_{r=0}^{l}\Gamma(l{+}1)B_{l}\Theta_{r}
W_{2r}G_{3,2}^{1,3}\left[\left. W_{2r} \right|{}_{ \varsigma_{2r}}^{1,0} \right],\nonumber
\tag{32}
\end{align}
and
\begin{align} 
C_{4}{=}&{1-P_{i}\over \Gamma(m_{1})\exp(K)}\sum_{l=0}^{k}B_{l} \nonumber\\
&\times G_{1,0;2,2;1,2}^{0,1;1,2;1,1}\left[\left. \begin{matrix} {-l}\\{-} \end{matrix} \right| \left. \begin{matrix} {1,1}\\{1,0} \end{matrix}\right| \left. \begin{matrix} {1}\\ {m_{1},0}\end{matrix} \right| {\overline{\gamma}_{RD}\over K+1},{\Omega_{1}\overline{\gamma}_{RD}\over m_{1}\left(K+1\right)} \right]\nonumber\\
&+{P_{i}\over \Gamma(m_{2})\exp(K)}\sum_{l=0}^{k}B_{l} \nonumber\\
&\times G_{1,0;2,2;1,2}^{0,1;1,2;1,1}\left[\left. \begin{matrix} {-l}\\{-} \end{matrix} \right| \left. \begin{matrix} {1,1}\\{1,0} \end{matrix}\right| \left. \begin{matrix} {1}\\ {m_{2},0}\end{matrix} \right| {\overline{\gamma}_{RD}\over K+1},{\Omega_{2}\overline{\gamma}_{RD}\over m_{2}\left(K+1\right)} \right], \nonumber
\tag{33}
\end{align}
where $\varsigma_{1r}=\left\{1{-}m_{1}{-}r,1,1\right\}$, $\varsigma_{2r}=\left\{1{-}m_{2}{-}r,1,1\right\}$, $\Theta_{r}={(K{+}1)^{r} \over \overline{\gamma}_{RD}^{r}\Gamma(r{+}1)}$,
$W_{1r}{=}\left({\Omega_{1}\overline{\gamma}_{RD}\over m_{1}\overline{\gamma}_{RD}{+}\Omega_{1}\left(K{+}1\right)}\right)^{m_{1}+r}$,
$W_{2r}{=}\left({\Omega_{2}\overline{\gamma}_{RD}\over m_{2}\overline{\gamma}_{RD}{+}\Omega_{2}\left(K{+}1\right)}\right)^{m_{2}+r}$. Furthermore, an upper bound for $C_{DF}$ can be formulated as \cite{31}
\begin{align}
C_{DF}=\frac{1}{2}\mathbb{E}\left\{\text{log}_{2}(1{+}\gamma) \right\}
\leq \frac{1}{2}\text{log}_{2}\left(1+\mathbb{E}(\gamma)\right),
\nonumber
\tag{34}
\end{align}
where the mean value $\mathbb{E}(\gamma)=\int_{0}^{\infty}\gamma f_{\gamma}(\gamma)$ can calculated as (35), shown at the top of this page.
\subsection{Fixed-Gain AF Relaying}
\subsubsection{OP}
For the AF relaying scheme, OP can be formulated as
\begin{align}
P_{\text{out}}^{AF}=\text{Pr}\left\{ \frac{\gamma_{\text{SR}}\gamma_{\text{RD}}}{C+\gamma_{\text{RD}}}<\gamma_{\text{th}}\right\}=F_{\gamma_{o}}(\gamma_{\text{th}}),
\nonumber
\tag{36}
\end{align}
where $F_{\gamma_{o}}(\gamma_{\text{th}})$ denotes the CDF of $\gamma_o$ when $\gamma{=}\gamma_{\text{th}}$.
\subsubsection{Average BER}
Assuming DBPSK scheme, the average BER for the considered cooperative PLC/RF system with fixed-gain AF relaying can be expressed as
$P_{\text{BER}}^{AF}{=}\frac{1}{2}\int_{0}^{\infty}{\rm exp}\left ( {-}\gamma  \right )F_{\gamma _{o}}\left ( \gamma  \right )d\gamma \nonumber $.
After some algebraic calculations, we can arrive at
the average BER in (37), shown at the top of this page.
\subsubsection{Average Channel Capacity}
By employing AF protocol, the ACC can be formulated as
\begin{align}
C_{AF}{=}\frac{1}{2}\mathbb{E}\left\{\text{log}_{2}(1{+}\gamma_{o}) \right\}
{=}\frac{1}{2\text{ln}(2)}\int_{0}^{\infty}\text{ln}(1{+}\gamma)f_{\gamma_{o}}(\gamma)d\gamma.
\nonumber
\tag{38}
\end{align}
Then, upon substituting (16) in (38) and using the integral from three Meijer $G$-functions [28, Eq. 07.34.21.0081.01],
$C_{AF}$ is given by (39), shown at the top of this page.

Furthermore, an upper bound for $C_{AF}$ can be represented as
\begin{align}
C_{AF}=\frac{1}{2}\mathbb{E}\left\{\text{log}_{2}(1{+}\gamma_{o}) \right\}
\leq \frac{1}{2}\text{log}_{2}\left(1+\mathbb{E}(\gamma_{o})\right),\nonumber
\tag{40}
\end{align}
where
\begin{align}
\mathbb{E}(\gamma_{o}){=}&{(1-P_{i})\Omega_{1}\exp({-}K)\over \Gamma(1{+}m_{1})}\sum_{l=0}^{k}\sum_{r=0}^{m_{1}}
\begin{pmatrix}m_{1}\\ r \end{pmatrix}\left({C(1{+}K)\over \overline{\gamma}_{RD}}\right)^{l{+}1} \nonumber\\
&\times B_{l} G_{2,1}^{1,2}\left[\left.\frac{C(1{+}K)}{\overline{\gamma}_{RD}} \right|\begin{matrix}r{-}m_{1}{-}l{-}1\\r{-}l{-}1,0\end{matrix}\right] \nonumber\\
&+{P_{i}\Omega_{2}\exp({-}K)\over \Gamma(1{+}m_{2})}\sum_{l=0}^{k}\sum_{r=0}^{m_{2}}\begin{pmatrix}m_{2}\\ r \end{pmatrix}\left({C(1{+}K)\over \overline{\gamma}_{RD}}\right)^{l{+}1}\nonumber\\
& \times B_{l}G_{2,1}^{1,2}\left[\left.\frac{C(1{+}K)}{\overline{\gamma}_{RD}} \right|\begin{matrix}r{-}m_{2}{-}l{-}1\\r{-}l{-}1,0\end{matrix}\right].\nonumber
\tag{41}
\end{align}
\section{Simulation and Numerical Results}
In this section, numerical examples are presented to illustrate the analytical
and asymptotic expressions developed in the previous section. Additionally,
Monte-Carlo simulation results corroborate the analytical analysis.
Unless otherwise specified, we set $m_1=m_{2}=8$, $P_{i}=0.2$, $\eta=5$, $\sigma_{SR}=0.23$, $K=6$ $\rm{dB}$, $C=1.2$, $G=3$, $\gamma_{th}=0$ $\rm{dB}$ and $\overline{\gamma}_{SR}=\overline{\gamma}_{RD}=\overline{\gamma}$.

In Fig. 2, OP versus $\overline{\gamma}$ is plotted for various values of $K$, and assuming
the considered dual-hop mixed PLC/RF system with the DF protocol. In addition, we set $K=2, 4, 6$ $\rm{dB}$.
\begin{figure}[t]
 \centering
 \includegraphics[scale=0.41]{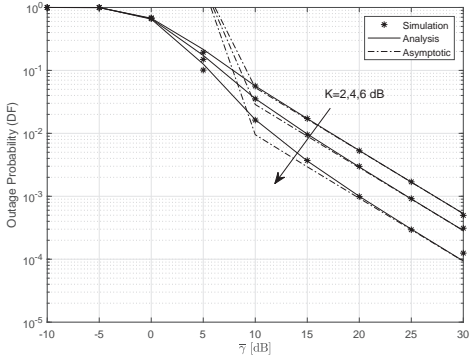}
 \caption{Outage probability for different values of $K$ (DF protocol). }
\end{figure}
\begin{figure}[t]
 \centering
 \includegraphics[scale=0.41]{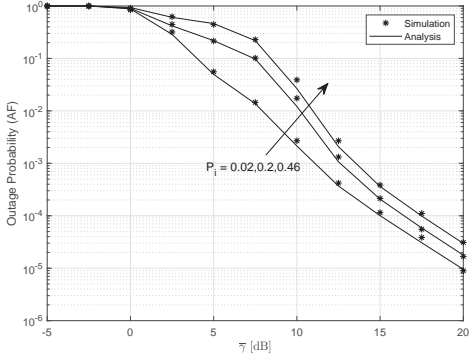}
 \caption{Outage probability for various values of $P_{i}$ (AF protocol). }
\end{figure}
It can be clearly noted that the system outage performance improves by increasing the values of $K$. The reason is that $K$
is the ratio of the powers of the LoS components to the powers of the scattered components. The higher the values of $K$,
the lower the signal attenuation caused by multipath effects. Furthermore, Fig. 2 shows that the asymptotic $P_{out}^{DF}$
expression converges to the exact $P_{out}^{DF}$ expression at high SNRs. In Fig. 3, we draw  $P_{out}^{AF}$
versus $\overline{\gamma}$ for various values of $P_{i}$ assuming the AF protocol. From Fig. 3, it can be observed that for low values
of $P_{i}$, $P_{out}^{AF}$ is highly degraded, resulting in a better performance. This is because $P_{i}$ represents the probability
of the arrival of the impulsive noise, and lower values of $P_{i}$ generates a weak effect of the IMN for the PLC/RF system.
Finally, Fig. 4 presents the outage performance of the dual-hop PLC/RF system with different values of threshold SNR $C_{th}$ under DF and AF relay protocols. From Fig. 4, it can be noted that the outage performance with the AF relay protocol is superior to the outage performance with the DF relay protocol, and the performance increases when the value $\gamma_{th}$ decreases.
\begin{figure}[t]
 \centering
 \includegraphics[scale=0.41]{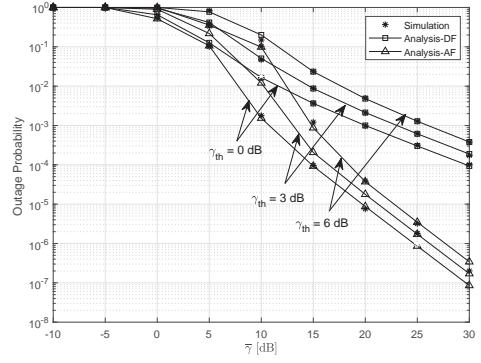}
 \caption{Outage probability for various values of $\gamma_{th}$ under differen relaying protocols. }
\end{figure}
\begin{figure}[t]
 \centering
 \includegraphics[scale=0.41]{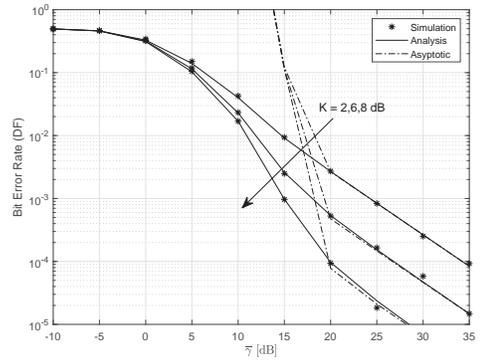}
 \caption{BER for different Rician factors $K$ (DF protocol). }
\end{figure}
\begin{figure}[t]
 \centering
 \includegraphics[scale=0.41]{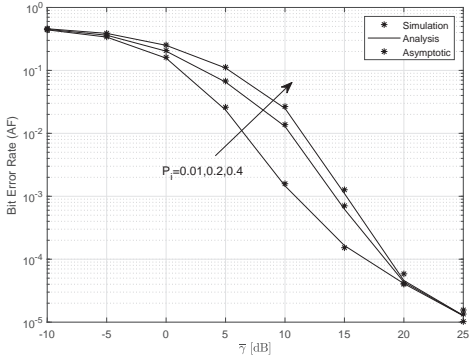}
 \caption{BER for various values of $P_{i}$ (AF protocol). }
\end{figure}
\begin{figure}[t]
 \centering
 \includegraphics[scale=0.41]{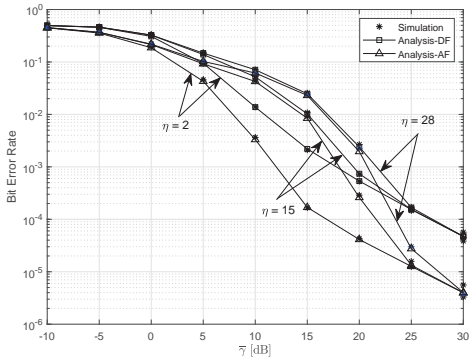}
 \caption{BER for various values of $\eta$ under different relaying protocols. }
\end{figure}

In Fig. 5, the average BER versus average SNR
$\overline{\gamma}$ is plotted under different values of $K$, and considering the DF relay protocol.
This figure demonstrates the average BER performance with higher values $K$ performs much better than the one with the lower values of $K$. The reason is because higher values of $K$ means more LoS components, which improves the system performance. Additionally, Fig. 5 also reveals the convergence between the asymptotic and exact $P_{BER}^{DF}$ expressions. The impact of the different values of $P_{i}$ is shown in Fig. 6. As can be seen, a decrease in the value of $P_{i}$ (the probability of impulse noise reaching the PLC channel) leads to a BER decrease, which results in a better system performance. The reason is that with the decrease of IMN arriving at the system, the severity of IMN component in the system decreases compared with the BGN sample, improving consequently the performance. In Fig. 7, the average BER is presented under differen values of $\eta$. It can be clearly seen that a decrease in $\eta$ leads to a lower average BER. Similar to the results obtained previously to outage probability, it is shown that the average BER of the AF protocol is significantly superior to the BER performance of the DF one.

Fig. 8 plots the ACC $C_{DF}$ versus $\overline{\gamma}$ for various values of $\eta$. It can be observed that $C_{DF}$ increases with the decrease of $\eta$. In addition, we plot the asymptotic $C_{DF}$ , which reveals the correctness of the asymptotic expression of $C_{DF}$. In Fig. 9, we plot exact average channel capacity $C_{AF}$ versus $\overline{\gamma}$ for various values of $P_{i}$. As can be seen, higher $C_{AF}$ can be obtained when the values of $P_{i}$ decrease. Also, it is shown the convergence between the asymptotic and exact $C_{AF}$ expressions.
\begin{figure}[t]
 \centering
 \includegraphics[scale=0.41]{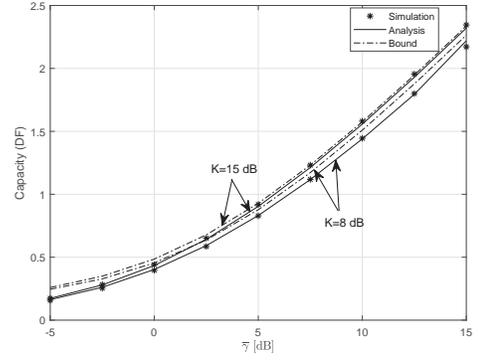}
 \caption{Ergodic channel capacity for various values of $K$ (DF protocol).}
\end{figure}
\begin{figure}[t]
 \centering
 \includegraphics[scale=0.41]{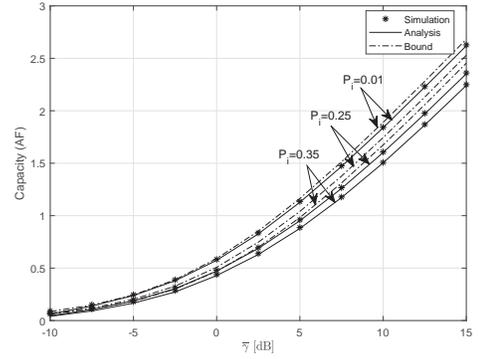}
 \caption{Ergodic channel capacity for different values of $P_{i}$ (AF protocol). }
\end{figure}
\section{Conclusions}
In this work, we studied the performance of the PLC/RF system under both DF and AF relay protocols. It was supposed that the PLC link is modeled by a LN distribution with IMN while the RF link follows the Rician distribution. Closed-form expressions for the OP, average BER, and ACC were derived. For AF relaying, analytical expressions for the CDF and PDF of the end-to-end SNR have been obtained in closed-form. For DF relaying, closed-form expressions for the asymptotic OP, average BER, and ACC were derived. The analytical expressions were validated with their corresponding simulation results. Additionally, it was studied the impact of impulsive noise and Rician factor on the overall system performance, and insightful discussions were drawn.
\appendices
\section{PDF AND CDF of The End-To-End SNR}
In this section, we derive the PDF and CDF of the end-to-end overall SNR $\gamma_{o}$. Similar methods can also be found in
\cite{27}. The PDF of $\gamma_{o}=\frac{\gamma_{\textrm{SR}}\gamma_{\textrm{RD}}}{C+\gamma_{\textrm{RD}}}$ can be
written as
\begin{align}
f_{\gamma_{o}}(\gamma)=&\frac{d}{d\gamma}\text{Pr}\left\{\frac{\gamma_{\textrm{SR}}\gamma_{\textrm{RD}}}{C+\gamma_{\textrm{RD}}}<\gamma\right\}\nonumber\\
=&\frac{d}{d\gamma}\int_{0}^{\infty}\text{Pr}\left \{ \frac{\gamma_{\textrm{RD}}x}{C+\gamma _{RD}}< \gamma  \right \}f_{\gamma _{\textrm{SR}}}\left ( x \right )dx\nonumber\\
=&\frac{d}{d\gamma }\left [ { \int_{0}^{\gamma }\text{Pr}\left \{ \gamma_{\textrm{RD}}\left ( x-\gamma  \right )< C\gamma  \right \}f_{\gamma _{\textrm{SR}}}\left ( x  \right )dx}\right. \nonumber \\
&\left.{+\int_{\gamma }^{\infty} \text{Pr}\left \{ \gamma _{\textrm{RD}}\left ( x-\gamma  \right )< C\gamma  \right \}f_{\gamma _{\textrm{SR}}}\left ( x  \right )dx} \right ]. \nonumber
\tag{42}
\end{align}
Due to $0<x<\gamma$, $\text{Pr}\left\{\gamma_{\textrm{RD}}(x-\gamma)<C\gamma\right\}=1$, (42) can be rewritten as
\begin{align}
f_{\gamma_{o}}(\gamma){=}&\frac{d}{dr}\left[{\int_{0}^{\gamma}f_{\gamma_{\textrm{SR}}}(x)dx} \right.\nonumber\\
&\left.{{+} \int_{\gamma }^{\infty} \text{Pr}\left \{ \gamma _{\textrm{RD}}< \frac{C\gamma}{x{-}\gamma}  \right \}f_{\gamma _{\textrm{SR}}}\left ( x  \right )dx }\right]\nonumber\\
=&f_{\gamma_{\textrm{SR}}}(\gamma)-\lim_{x\rightarrow \gamma ^{+}}\text{Pr}\left\{\gamma_{\textrm{RD}}<\frac{C\gamma}{x-\gamma}\right\}f_{\gamma_{SR}}(\gamma)\nonumber\\
&+\int_{\gamma}^{\infty}f_{\gamma_{\textrm{RD}}}\left(\frac{C\gamma}{x-\gamma}\right)\frac{cx}{(x-\gamma)^2}f_{\gamma_{\textrm{SR}}}(x)dx.\nonumber
\tag{43}
\end{align}
Since $\lim_{x\rightarrow \gamma ^{+}}\text{Pr}\left\{\gamma_{\textrm{RD}}<\frac{C\gamma}{x-\gamma}\right\}=1$, one can obtain
\begin{align}
f_{\gamma_{o}}(\gamma)= \int_{\gamma}^{\infty}f_{\gamma_{\textrm{RD}}}(\left(\frac{C\gamma}{x-\gamma}\right)\frac{cx}{(x-\gamma)^2}f_{\gamma_{\textrm{SR}}}(x)dx.\nonumber
\tag{44}
\end{align}
By substituting $t=x-\gamma$ in (44) and applying (6), (14),  $f_{\gamma_{o}}(\gamma)$ can be formulated as
\begin{align}
&f_{\gamma_{o}}(\gamma){=}{\exp(-{m_{1}\over \Omega_{1}}\gamma)\left( { m_{1}\over \Omega_{1} }\right)^{m_{1}}\over \Gamma(m_{1})\exp(K)}\sum_{l{=}0}^{k}
\left({C(1+K)\over \overline{\gamma}_{RD}}\right)^{l+1}B_{l}\gamma^{l}\Phi_{1}\nonumber\\
&{+}{\exp(-{m_{2}\over \Omega_{2}}\gamma)\left( { m_{2}\over \Omega_{2} }\right)^{m_{2}}\over \Gamma(m_{2})\exp(K)}\sum_{l{=}0}^{k}
\left({C(1+K)\over \overline{\gamma}_{RD}}\right)^{l+1}B_{l}\gamma^{l}\Phi_{2},
\tag{45}
\end{align}
where $\Phi_{\rho}$, $\rho=1,2$ are expressed as
\begin{align}
\Phi_{\rho}{=}&\int_{0}^{\infty}\left(1{+}\frac{\gamma}{t}\right)^{m_{\rho}}t^{m_{\rho}{-}l{-}2}\nonumber\\
&\times\exp\left({-}\frac{m_{\rho}t}{\Omega_{\rho}}\right)\exp({-}{C(K{+}1)\over t\overline{\gamma}_{RD}}\gamma)dt.\nonumber
\tag{46}
\end{align}
Using $\exp({-}bz){=}G_{0,1}^{1,0}\left[\left.bz \right|{}_{0}^{-}\right]$, the integral formula [28, Eq. (07.34.21.0088.01)] and the expanding expression of $(1+\gamma/t)^{m_{\rho}}$, $\rho=1,2$ \cite{27}, (16) is obtained.

From (43), the CDF of $\gamma_{o}$ can be written as
\begin{align}
F_{\gamma_{o}}(\gamma){=}F_{\gamma_{SR}}(\gamma){+}\int_{\gamma}^{\infty}F_{\gamma_{RD}}({C\gamma\over x-\gamma})f_{\gamma_{SR}}(x)dx.\nonumber
\tag{47}
\end{align}
To simplify the operation, we use the new expression of $F_{\gamma_{RD}}(\gamma){=}\sum_{l=0}^{k}{k^{1{-}2l}K^l\Gamma(k+l) \over \Gamma(k{-}l+1)\Gamma^2(k+1)\exp(K)}\Upsilon(l{+}1,{ (K{+}1)\gamma \over \overline{\gamma}_{RD}})$ obtained from the integration of (14) to (47). Therefore, the CDF of $\gamma_{o}$ is represented by (17).

\def\baselinestretch{0.98}

\end{document}